\UseRawInputEncoding
\documentclass[showpacs,superscriptaddress,twocolumn]{revtex4-2}
\usepackage{bm,graphicx,dcolumn,epstopdf,epsf,latexsym,mathbbol,dcolumn,amssymb,amsmath,color,slashed,mathrsfs,mathcomp,simplewick}

\usepackage{multirow}
\usepackage{tabularx}
\usepackage{makecell}
\usepackage{color, soul}
\usepackage{hyperref}
\usepackage{url}

\usepackage{amssymb}
\usepackage{amsmath}
\usepackage{amsfonts}
\usepackage{slashed}
\usepackage{graphicx}
\usepackage{relsize}
\usepackage{color}
\usepackage{extarrows}
\usepackage{cancel}
\usepackage{placeins}
\usepackage{float}
\usepackage{subcaption}
\usepackage{caption}
\usepackage{here}
\usepackage[normalem]{ulem}

\begin{document}

\title{
Thermodynamics of Einstein-Geometric Proca AdS Compact Objects}
\author{Asalkhon Alimova}
\email{asalxon2197@gmail.com}

\affiliation{Tashkent State Technical University, Tashkent 100095, Uzbekistan}
\affiliation{National Research University TIIAME, Kori Niyoziy 39, Tashkent 100000, Uzbekistan}

\author{Elham Ghorani}
\email{elham.ghorani@sabanciuniv.edu}
\affiliation{Faculty of Engineering and Natural Sciences, Sabanc{\i} University, 34956 Tuzla, Istanbul, Turkey}

\author{Beyhan Puli{\c c}e}
\email{beyhan.pulice@sabanciuniv.edu}
\affiliation{Faculty of Engineering and Natural Sciences, Sabanc{\i} University, 34956 Tuzla, Istanbul, Turkey}
\affiliation{Astrophysics Research Center,
The Open University of Israel, Raanana 4353701, Israel}

\author{Farruh Atamurotov}
\email{atamurotov@yahoo.com}
\affiliation{Kimyo International University in Tashkent, Shota Rustaveli str. 156, Tashkent 100121, Uzbekistan}
\affiliation{University of Tashkent for Applied Sciences, Str. Gavhar 1, Tashkent 100149, Uzbekistan}
\affiliation{Research Center of Astrophysics and Cosmology, Khazar University, 41 Mehseti Street, Baku AZ1096, Azerbaijan}

\author{Ahmadjon Abdujabbarov}
\email{ahmadjon@astrin.uz}

\affiliation{New Uzbekistan University, Movarounnahr street 1, Tashkent 100000, Uzbekistan}
\affiliation{School of Physics, Harbin Institute of Technology, Harbin 150001, People’s Republic of China}

\begin{abstract}
In this study, we explore metric-Palatini gravity extended by the antisymmetric component of the affine curvature. This gravitational theory results in general relativity plus a geometric Proca field. Building on our previous work, where we constructed its static spherically symmetric solutions in the Anti-de Sitter (AdS) background [Eur. Phys. J. C83 (2023) 4, 318], we conduct a comprehensive analysis of the system's thermodynamics. We examine the thermodynamic properties of the Einstein-Geometric Proca AdS compact objects, focusing on the Hawking temperature, enthalpy, heat capacity, entropy, and Gibbs free energy. Particular attention is given to the dependence of the Hawking temperature, enthalpy, and heat capacity on the uniform potential $q_1$ and the electromagnetic-type charge $q_2$. Through numerical analysis, we compute the entropy and Gibbs free energy and investigate how these quantities vary with the model parameters.

\end{abstract}

\maketitle

\section{Introduction \label{sec1}}

The thermodynamics of black holes, first formulated by Bekenstein and Hawking, establish a connection between gravity, quantum mechanics, and entropy. The Bekenstein-Hawking entropy is given by $S=kc^{3}A/4G\hbar$, linking the black hole entropy to horizon geometry \cite{1973PhRvD...7.2333B,1975CMaPh..43..199H}. Hawking radiation, a quantum effect near the event horizon of the black hole, causes black holes to emit thermal radiation energy and gradually lose mass, potentially leading to complete evaporation \cite{Hawking1974Nature}. These principles mirror classical thermodynamics, with the black hole temperature related to surface gravity and entropy to the horizon area in~\cite{1994qftc.book.....W}. In modified gravity theories such as $f(R)$ gravity, Gauss-Bonnet gravity, and scalar-tensor theories, black hole thermodynamics deviates from these classical laws due to additional curvature corrections, extra fields, or higher-dimensional effects in Refs~.\cite{2002PhRvD..65h4019C,2011PhR...505...59N}. Extensive research has been conducted to establish a precise framework for interpreting the thermodynamic properties of black holes in various gravities in \cite{BHT2,BHT4,BHT5,2023NuPhB.99616363R,2023AnPhy.45369326D,2023NuPhB.99016180J,2022EPJC...82..756D,2024NewA..10502106Y,2025EPJC...85..229S,2025ChPhC..49g5105R}.

Current research in astrophysics, gravitation, and cosmology is centered on a fundamental inquiry. Is general relativity (GR) the only possible theory of gravity? Answering this requires an in-depth exploration of alternative theories that extend GR in physically meaningful ways. One such extension involves non-Riemannian geometries in which the metric and connection are treated as independent geometric quantities \cite{schroedinger,mag1,Vitagliano2011}.

A fundamental extension in this framework is metric-Palatini gravity \cite{harko2012,Capozziello2013a,Capozziello2015}, which has been extensively explored in various domains, including dark matter dynamics \cite{Capozziello2012a}, the formation of wormholes \cite{Capozziello2012b}, and cosmological applications \cite{Capozziello2012c}. A notable variant of this theory arises when the non-metricity tensor generates a geometric $Z^\prime$ field, defining a distinct class of models \cite{Demir2020}. This formulation has been the subject of studies on gravitational waves \cite{Shaaban} and black hole properties in both Schwarzschild \cite{dp-yeni} and AdS \cite{AdS-Proca-1} spacetimes. Expanding on our earlier investigation \cite{AdS-Proca-1}, this work is dedicated to analyzing Einstein-Geometric Proca AdS objects in greater depth.

The metric-Palatini gravity framework has been extensively analyzed in \cite{dp-yeni,AdS-Proca-1}, and here we provide a concise overview of its key aspects. This theory is defined by a metric $g_{\mu\nu}$ and a torsion-free affine connection $\Gamma^\lambda_{\mu\nu}$, which remains independent of the Levi-Civita connection associated with the metric.

Metric-Palatini gravity has broad implications in various domains, including symmergent gravity, which restores gauge symmetry \cite{Demir2019, Demir2021, Demir2023}, its role in natural inflation \cite{bauer-demir1, bauer-demir2}, and its astrophysical and cosmological significance \cite{Palatini-f(R), 2024PDU....4601577K,2025PDU....4701799K}. Furthermore, higher-curvature modifications within this framework have been explored in relation to fundamental physics \cite{Vitagliano2011, Vitagliano2013, Demir2020}.

Extending the Palatini formulation further, one can introduce a term of the form 
${\mathbb{R}}_{[\mu\nu]}(\Gamma) {\mathbb{R}}^{[\mu\nu]}(\Gamma)$ where ${\mathbb{R}}_{[\mu\nu]}(\Gamma)$ represents the antisymmetric component of the affine Ricci tensor ${\mathbb{R}}_{\mu\nu}(\Gamma)$. This modification is particularly significant, as it leads to a formulation that encompasses general relativity (GR) along with a massive geometric vector field $Q_{\mu}$ \cite{Vitagliano2010,Demir2020}. This vector field, known as the geometric Proca field, is defined as $Q_\mu \equiv \frac{1}{4} Q _{\mu \nu}^{~~~\nu}$ where the non-metricity tensor is given by $Q_{\lambda\mu\nu}\equiv -{}^\Gamma \nabla_\lambda g_{\mu\nu}$. This construction naturally emerges within the Palatini framework and has been explored in various studies \cite{Demir2020, Buchdahl1979, Tucker1996, Obukhov1997, Vitagliano2010}.

In the absence of torsion, the non-metricity vector becomes the sole source of deviations from GR. The geometric Proca field naturally emerges as a direct consequence of metric-incompatible symmetric connections (which are torsion-free) rather than putting by hand. Unlike a gauge field, it represents a fundamentally geometric massive vector field \cite{Demir2020}, characterized by specific coupling interactions with quarks and leptons \cite{dp-yeni}. This Palatini framework can be further extended by incorporating both the metrical curvature $R_{\mu\nu}({}^g\Gamma)$ and the affine curvature ${\mathbb{R}}_{\mu\nu}(\Gamma)$ into the action, allowing for a more comprehensive formulation of the theory.

Apart from the quadratic term ${\mathbb{R}}_{[\mu\nu]}(\Gamma) {\mathbb{R}}^{[\mu\nu]}(\Gamma)$, which gives rise to the geometric Proca field, the combined metric-affine approach reduces to metric-Palatini gravity \cite{harko2012, Capozziello2013a, Capozziello2015}. The gravitational theory investigated in this work is essentially an extension of metric-Palatini gravity, augmented by the inclusion of the invariant ${\mathbb{R}}_{[\mu\nu]}(\Gamma) {\mathbb{R}}^{[\mu\nu]}(\Gamma)$ and a negative cosmological constant (CC) \cite{AdS-Proca-1, AdS-Proca-2}. As demonstrated in \cite{dp-yeni}, the presence of the geometric Proca field $Q_{\mu}$ requires the inclusion of a CC for the existence of static spherically symmetric solutions. We refer to this framework as extended metric-Palatini gravity (EMPG). The corresponding action follows a schematic structure, as described in \cite{AdS-Proca-1}. 
 \begin{eqnarray}
\label{EMPG}
S[g,\Gamma]&=&\!\!\int\!\! d^4x \sqrt{-g} \Bigg \{\!``g^{\mu\nu}{R}_{\mu\nu}\left({}^g\Gamma\right)"+ ``g^{\mu\nu}{\mathbb{R}}_{\mu\nu}\left(\Gamma\right)"
 \nonumber \\ &&+ ``{\mathbb{R}}_{[\mu\nu]}(\Gamma) 
 {\mathbb{R}}^{[\mu\nu]}(\Gamma)" + ``{\rm CC}" \!\Bigg \} .
\end{eqnarray} 

This framework constitutes an Einstein-Geometric Proca-Anti de Sitter (AdS) gravity theory, distinguished from traditional Einstein-Proca models by its purely geometric foundation. Unlike conventional Einstein-Proca systems, which have been extensively studied in the literature for various purposes, such as exploring Reissner-Nordström-type spherically symmetric vacuum solutions \cite{Tresguerres1995a, Tucker1995, Vlachynsky1996, Macias1999}, investigating the role of the Proca field \cite{Bekenstein1971, Bekenstein1972, Adler1978}, deriving static spherically symmetric solutions \cite{Frolov1978, Gottlieb1984, Leaute1985}, and analyzing the structure of the horizon radius \cite{Ayon1999, Obukhov1999, Toussaint2000}, our formulation emerges naturally from the metric-affine approach. 

The objective of this work is to build on our previous study \cite{AdS-Proca-1} by conducting a detailed analysis of the thermodynamics of compact objects within the extended metric-Palatini gravity (EMPG) framework.

The remainder of this article is organized as follows: We analyze  static spherically-symmetric solutions in EMPG model in Section \ref{sect:2}. Section \ref{sec3}
 focuses on the thermodynamical properties of Einstein-Geometric Proca compact objects. The behavior of Hawking temperature has been examined in Section \ref{sec3A}. Enhalpy, heat capacity, entropy and Gibbs energy have been discussed in Sections \ref{sec3B},\ref{sec3C}, \ref{sec3D},\ref{sec3E}, respectively. Finally, we summarize our findings in Section \ref{sec4}.

\section{Static Spherically-Symmetric Solutions in EMPG Model\label{sect:2}}

This section presents a brief discussion of the EMPG model, with much of the content drawn from our previous work \cite{AdS-Proca-1}. Here, we summarize the key findings. The EMPG action is provided in \cite{Demir2020, dp-yeni, AdS-Proca-1, AdS-Proca-2}.
 \begin{eqnarray}
S[g,\Gamma]&=&\int d^4x \sqrt{-g} \Bigg \{
\frac{M^2}{2} {R}\left(g\right) + \frac{{\overline{M}}^2}{2} {\mathbb{R}}\left(g,\Gamma\right) 
 \nonumber \\ &&+ \xi {\overline{\mathbb{R}}}_{\mu\nu}\left(\Gamma\right) {\overline{\mathbb{R}}}^{\mu\nu}\left(\Gamma\right) -V_0 + 
{\mathcal{L}}_{m}({}^g\Gamma,\psi) \Bigg \}, 
\label{mag-action}
\end{eqnarray}
where the affine curvatures in this action follow from the affine Riemann curvature
\begin{eqnarray}
\label{affine-Riemann}
{\mathbb{R}}^\mu_{\alpha\nu\beta}\left(\Gamma\right) = \partial_\nu \Gamma^\mu_{\beta\alpha} - \partial_\beta \Gamma^\mu_{\nu\alpha} + \Gamma^\mu_{\nu\lambda} \Gamma^\lambda_{\beta\alpha} -\Gamma^\mu_{\beta\lambda} \Gamma^\lambda_{\nu\alpha}, 
\end{eqnarray}
 with ${\mathbb{R}}^\mu_{\alpha\nu\beta}\left(\Gamma\right)=-{\mathbb{R}}^\mu_{\alpha\beta\nu}\left(\Gamma\right)$. Its contractions give rise to two distinct affine Ricci tensors: the canonical tensor ${\mathbb{R}}_{\mu\nu}\left(\Gamma\right) = {\mathbb{R}}^\lambda_{\mu\lambda\nu}\left(\Gamma\right)$, and the antisymmetric Ricci tensor ${\overline{\mathbb{R}}_{\mu\nu}}\left(\Gamma\right) = {\mathbb{R}}^\lambda_{\lambda\mu\nu}\left(\Gamma\right) = {\mathbb{R}}_{[\mu\nu]}\left(\Gamma\right)$.
In metrical geometry, the latter vanishes identically, i.e., ${\overline{\mathbb{R}}_{\mu\nu}}\left({}^g\Gamma\right) = 0$.
The term proportional to $M^2$ in the action (\ref{mag-action}) corresponds to the Einstein-Hilbert term in GR. The term proportional to ${\overline{M}}^2$ corresponds to the linear case of metric-Palatini gravity. The third term, proportional to $\xi$, gives the extension of metric-Palatini gravity with the antisymmetric part of the affine Ricci curvature \cite{Demir2020,dp-yeni}. In the last two terms, we separate the vacuum energy density $V_0$ from the Lagrangian of matter ${\mathcal{L}}_{m}({}^g\Gamma,\psi) $ that describes the dynamics of the matter fields $\psi$. 

The torsion-free affine connection can always be decomposed as 
\begin{align}\Gamma^\lambda_{\mu\nu}= {}^g\Gamma^\lambda_{\mu\nu} +  \frac{1}{2} g^{\lambda \rho} ( Q_{\mu \nu \rho } + Q_{\nu \mu \rho } - Q_{\rho \mu \nu} ),
\label{fark-connection}
\end{align}
with the Levi-Civita connection ${}^g\Gamma^\lambda_{\mu\nu}$ and $Q_{\lambda \mu \nu} = - {}^\Gamma \nabla_{\lambda} g_{\mu \nu}$ the non-metricity tensor. Applying this decomposition to the metric-Palatini action (\ref{mag-action}) results in the reduced action \cite{Demir2020, dp-yeni, AdS-Proca-1}
\begin{eqnarray}\label{action-reduced}
S[g,Y,\psi] &=& \int d^4 x \sqrt{-g} \Bigg\{\frac{1}{16 \pi G_N} R(g)-V_0 - \frac{1}{4} Y_{\mu \nu} Y^{\mu \nu}  \nonumber \\ &&- \frac{1}{2} M_Y^2 Y_{\mu} Y^{\mu}+{\mathcal{L}}_{m} (g,{}^g \Gamma,\psi) \Bigg \}, 
\end{eqnarray}
in which $Q_\mu = Q _{\mu \nu}^{\nu}/4$ is the non-metricity vector,  $Y_\mu=2 \sqrt{\xi} Q_\mu$ is the canonical geometric Proca field, $G_N=8\pi/(M^2 + \overline{M}^2)$ is Newton's gravitational constant, and $M_Y^2 = \frac{3 \overline{M}^2 }{2 \xi}$ is the squared mass of the $Y_\mu$. For the purposes of this analysis, it is convenient to express the reduced action (\ref{action-reduced}) in geometrical units as
 \begin{eqnarray}\label{action-EGP}
     S[g,Y] &=& \int d^4 x \sqrt{-g} \frac{1}{2 \kappa}  \nonumber \\ && \times \Bigg\{R(g) - 2 \Lambda  -  M_Y^2 \hat{Y}_{\mu} \hat{Y}^{\mu}
- \frac{1}{2} \hat{Y}_{\mu \nu} \hat{Y}^{\mu \nu} \Bigg\} ,
 \end{eqnarray}
in which $\kappa = 8 \pi G_N$, $\Lambda=8\pi G_N V_0$ is the CC, and $\hat{Y}_\mu \equiv \sqrt{\kappa} Y_{\mu}$ is the canonical dimensionless Proca field. From the given action, the equations of motion for the metric tensor $g_{\mu\nu}$ and the field $\hat{Y}_\mu$
can be derived using the principle of least action,
\begin{eqnarray}\label{Einstein-eqns}\nonumber
   R_{\mu \nu} - \Lambda g_{\mu \nu} - \hat{Y}_{\alpha\mu} \hat{Y}^{\alpha}_{\;\;\;\nu}  + \frac{1}{4} \hat{Y}_{\alpha \beta} \hat{Y}^{\alpha \beta} g_{\mu \nu} - M_Y^2 \hat{Y}_{\mu} \hat{Y}_{\nu} = 0, 
\end{eqnarray}
and 
\begin{align}
\nabla_\mu \hat{Y}^{\mu \nu} - M^2_Y \hat{Y}^\nu = 0.
\label{eom-Y}
\end{align}
These equations have been thoroughly investigated in \cite{dp-yeni} and \cite{AdS-Proca-1}, aiming to obtain black hole solutions for 
$\Lambda=0$ and $\Lambda<0$, respectively. In pursuit of a general static spherically symmetric solution, the following ansatz is employed
\begin{align}
g_{\mu \nu} = \text{diag}(-h(r),\frac{1}{f(r)} ,r^2, r^2 \sin ^2 \theta), 
\label{metric-sss}
\end{align}
and the vector field satisfying the equation of motion(\ref{eom-Y}), can be considered as purely time-like
\begin{align}
\hat{Y}_\mu = \hat{\phi}(r) \delta_\mu^0 
\label{proca-sss}
\end{align} 
lead to the solution of the Proca field
\begin{equation}
\hat{\phi}(\hat{r})=\frac{q_1}{\hat{r}^{\frac{1-\sigma}{2}}} + \frac{q_2}{\hat{r}^{\frac{1+\sigma}{2}}} , 
\label{psi}
\end{equation}
where $\sigma=\sqrt{1 + 4 \hat{M}_Y^2 l^2}$ in which $l$ stands for the AdS radius, and where we define the following dimensionless quantities:
\begin{align}
\hat{r} := \kappa^{-1/2} r,~ 
{\hat M}_Y^2 := \kappa M_Y^2~ .
\end{align}
The Breitenlohner-Freedman mass bound \cite{EProca} lets the range $0 \leq \sigma < 1$. This configuration avoids tachyonic runaway instabilities in the AdS background, with $q_1$ and $q_2$ corresponding to a uniform potential and an electromagnetic type charge, respectively. Consequently, the metric components $f$ and $h$ follow the form given in \cite{AdS-Proca-1}, in association with the geometric Proca solution (\ref{psi})

\begin{figure*}[t]
    \begin{center}
    \includegraphics[scale=0.7]{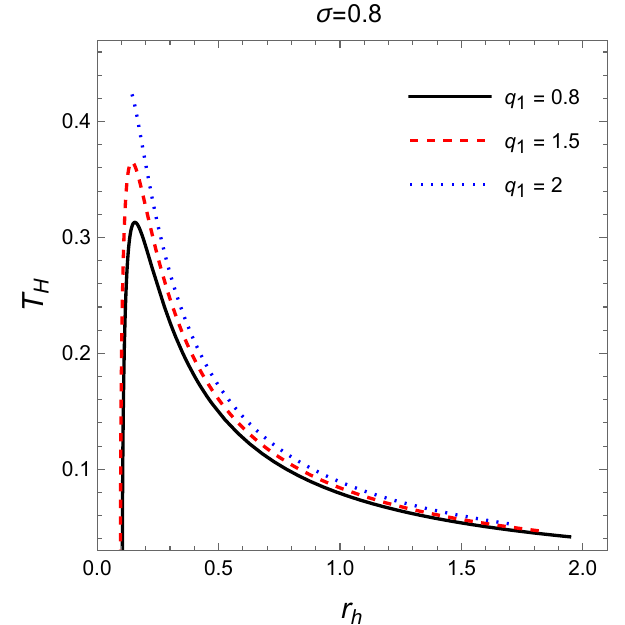}
\includegraphics[scale=0.7]{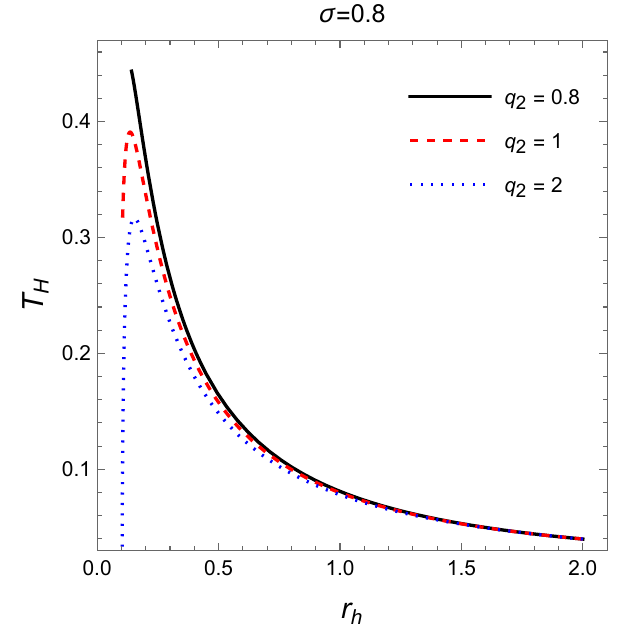}
 \caption{Hawking temperature  $T_H$ with respect to horizon radius $r_h$ for fixed values of $q_1$ and $q_2$.}
    \label{HTr}
    \end{center}
\end{figure*}
\begin{figure*}[t]
\includegraphics[scale=0.7]{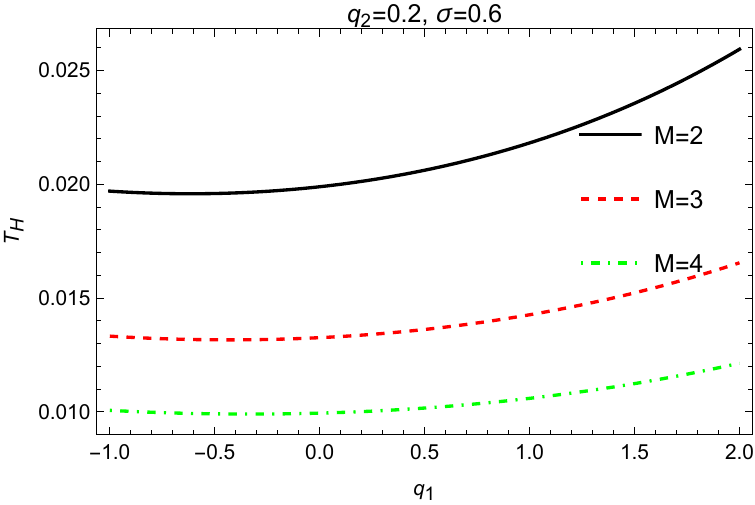} 
 \includegraphics[scale=0.7]{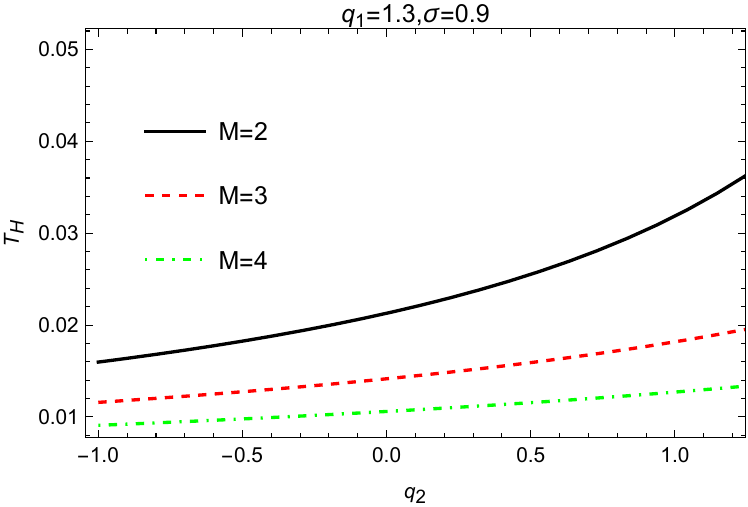}
  \caption{Hawking temperature $T_H$ with respect to $q_1$ (on the left) and $q_2$ (on the right) for fixed values mass.}
   \label{HT}
 \end{figure*}
 
\begin{align}
\label{metric-funcs-param}
f(\hat{r}) &= \hat{r}^2l^{-2} + 1 + \frac{n_1}{\hat{r}^{1 - \sigma}} + \frac{n_2}{\hat{r}}\ ,\nonumber\\
h(\hat{r}) &= \hat{r}^2l^{-2} + 1 +\frac{m_1}{\hat{r}^{1 - \sigma}}+\frac{m_2}{\hat{r}}\ ,
\end{align}
in which
\begin{align}
\label{solution}
&n_1 = \frac{1 - \sigma}{4} q_1^2\ , \quad m_1 = \frac{1 - \sigma}{3 - \sigma} q_1^2\ , \nonumber \\
&n_2 = m_2 - \frac{(1 - \sigma)(1 + \sigma)}{6} q_1 q_2 \ , 
\end{align}
Setting $q_1=0$ yields the AdS-Schwarzschild solution. The ADM mass of the resulting compact object is given by the expression in \cite{AdS-Proca-1}
\begin{eqnarray}
\label{mass}
    M_{ADM}=\frac{1}{2}\left(q_1 q_2 \left[\gamma  \sigma +\frac{1}{3} (1-\sigma ) (\sigma +4)\right]-m_2\right). \ \ \ 
\end{eqnarray}
In this expression, $\gamma$ represents the surface term coefficient of the geometric Proca field under the normalization $M_{ADM}=1$. Comprehensive analyses of the physical characteristics of the Einstein-geometric Proca AdS solution, including the dependence of the horizon radius on the model parameters and its singularity structure, can be found in \cite{AdS-Proca-1}.
\begin{figure*}[t]
    \begin{center}
    \includegraphics[scale=0.7]{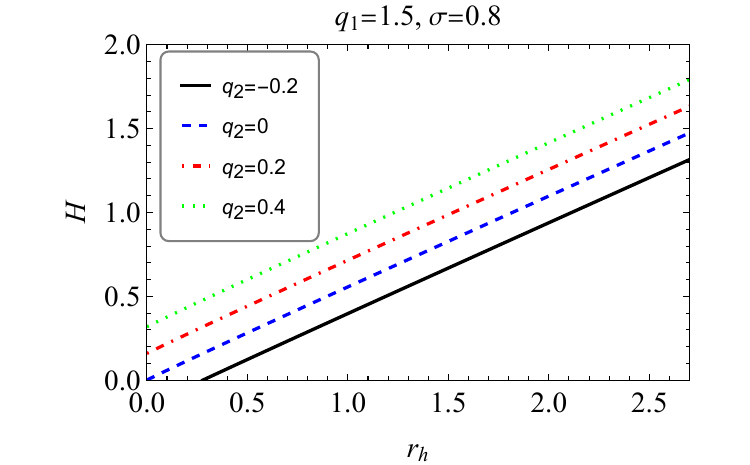}
    \includegraphics[scale=0.7]{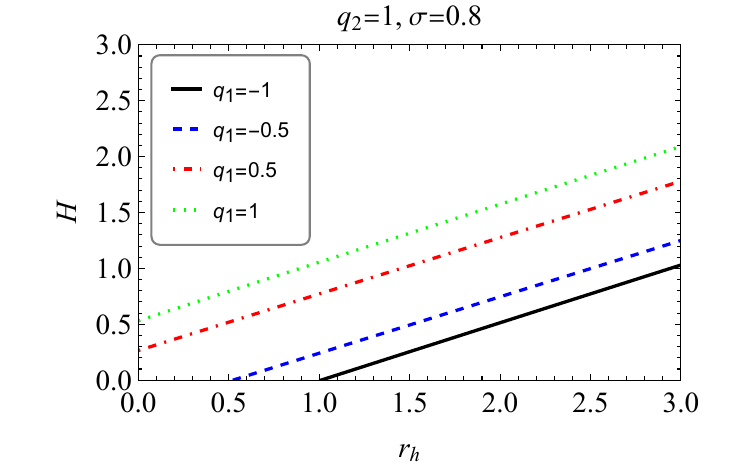}

    \caption{Enthalpy $H$ with respect to horizon radius $r_h$ for fixed values of $q_1$ and $q_2$.}
    \label{H}
    \end{center}
\end{figure*}
\begin{figure*}[t]
    \begin{center}
    \includegraphics[scale=0.7]{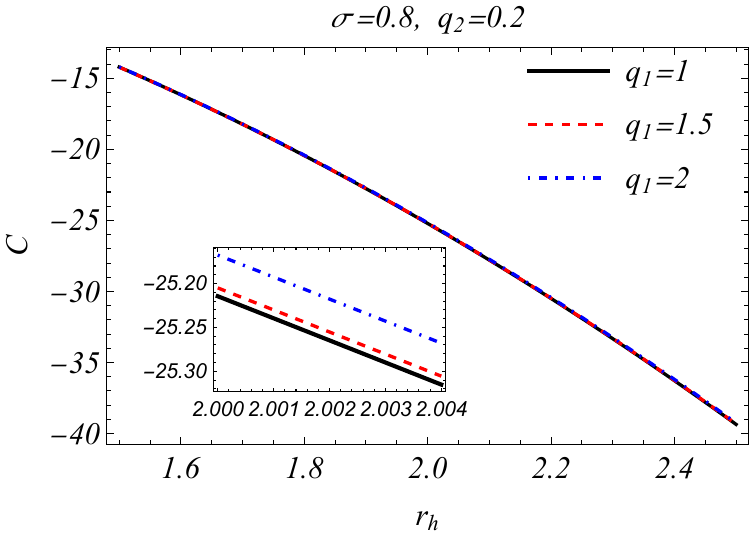}
    \includegraphics[scale=0.7]{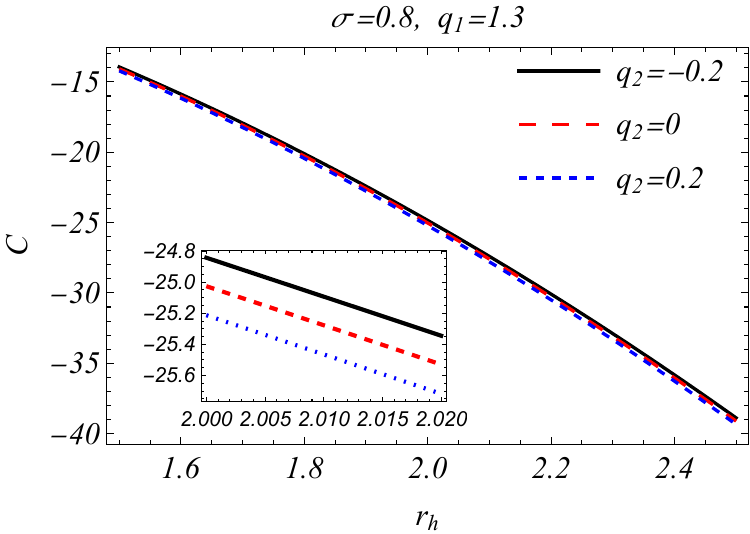}

    \caption{Heat capacity $C$ with respect to horizon radius $r_h$ for fixed values of $q_1$ and $q_2$.}
    \label{HC}
    \end{center}
\end{figure*}
\begin{figure*}[t]
    \begin{center}
    
    \includegraphics[scale=0.8]{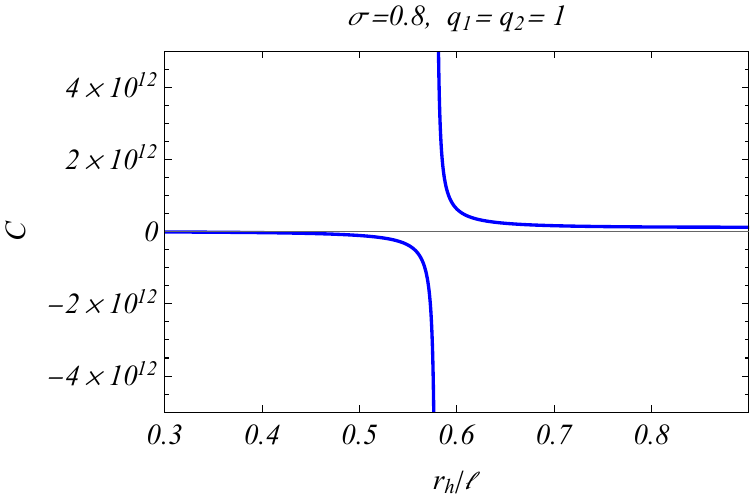}
    
     \caption{Heat capacity of Einstein-Geometric Proca AdS compact object as a function of the normalized event horizon $r_h/\ell$.}
    \label{phase}
    \end{center}
\end{figure*}

\section{Thermodynamical Properties in EMPG Model \label{sec3}}
The investigation of black hole thermodynamics revealed a fundamental connection among gravity, thermodynamics, and quantum theory. This understanding has been attained by classical and semi-classical analysis, greatly improving our understanding of quantum processes in strong gravitational fields \cite{Wald2001}. Progress in quantum field theory on curved surfaces has demonstrated an immediate link between surface gravity and temperature~\cite{Unruh1976}, as well as between the area of the event horizon and entropy~\cite{Bekenstein73}.
Subsequently, we investigate the thermodynamic characteristics of Einstein-Geometric Proca AdS Compact Objects. To facilitate calculations, we focus on the equatorial plane. In this section, we study the thermodynamic properties in the EMPG presented above, such as temperature, entropy, and heat capacity.

\subsection{Hawking Temperature \label{sec3A}}

In 1974, Hawking discovered that the physical temperature of a black hole is not absolute zero. Due to quantum particle creation effects, a black hole emits all types of particles to infinity with a perfect black body spectrum at temperature \cite{Wald2001}
\begin{equation}
T=\frac{\kappa}{2\pi}, 
\label{tha}
\end{equation}
where $\kappa$ is the surface gravity of BH, and we can calculate it with the following expression
\begin{equation}
\kappa= \frac{\sqrt{h^\prime(r) f^\prime(r)}}{2 } |_{r=r_{h}}. 
\label{kappa}
\end{equation}

After substituting Eqs. (\ref{tha})-(\ref{kappa})  Hawking temperature of a compact object with static spherically symmetric metric (\ref{metric-sss}) is expressed by \begin{equation}
    T_H =  \frac{\sqrt{h^\prime(r) f^\prime(r)}}{4 \pi } |_{r=r_{h}} , 
\end{equation}
which gives the following expression for the Hawking temperature using the metric solutions (\ref{metric-funcs-param}).
\begin{widetext}
\begin{equation}
\label{Eq-temp}
T_H = \frac{1}{16 l^2 r_h^2 \sqrt{3} \pi} 
\sqrt{\frac{
\left[12 r_h^3 + 4l^2r_h - q_1^2l^2 r_h^\sigma (\sigma-1) \sigma\right]
\left[3 l^2 q_1^2 r_h^\sigma (1 -4\sigma +3\sigma^2) + 
2 (\sigma-3) \left(18 r_h^3 +  6r_hl^2 + q_1 q_2l^2 (\sigma^2-1)\right)\right]
}{
(\sigma-3)
}} . 
\end{equation}
\end{widetext}

The event horizon of the black hole for the specified metric (\ref{metric-funcs-param}) is determined by the condition $f(\hat{r}) = 0$. If we consider $q_1=0$, the Hawking temperature of the Schwarzschild black hole $T_0 = 1/(8\pi M)$ is recovered. Fig.~\ref{HTr} illustrates the radial dependence of Hawking temperature on various values of $q_1$ and $q_2$.  It is obvious that $T_H$ decreases monotonically with increasing radial distance. Higher values of $q_1$ are associated with greater $T_H$, while higher values of $q_2$ correspond to lower $T_H$. Fig. \eqref{HT} shows that the Hawking temperature increases monotonically with the charge parameters while larger black hole masses correspond to lower temperatures. This behavior is consistent with the fact that massive black holes are colder, whereas charge contributions enhance the surface gravity.
\vspace{0.5cm}
\subsection{Enthalpy \label{sec3B}}

\vspace{0.3cm}
In black hole thermodynamics, the mass $M$ of a black hole is interpreted as its enthalpy $H$, not its internal energy, and can be found by the condition $f(\hat{r_h})=0$.
\begin{widetext}
\begin{equation}
\label{entalpy}
H=\frac{
  q_1 q_2 l^2(14 -12 \sigma+12  \gamma \sigma - 2 \sigma^2)+12  r_h(l^2+r_h^2) + 3 l^2 q_1^2 r_h^\sigma(1- 
    \sigma) 
}{24 l^2}. 
\end{equation}
\end{widetext}

\begin{figure*}[t]
 \begin{center}
\includegraphics[scale=0.7]{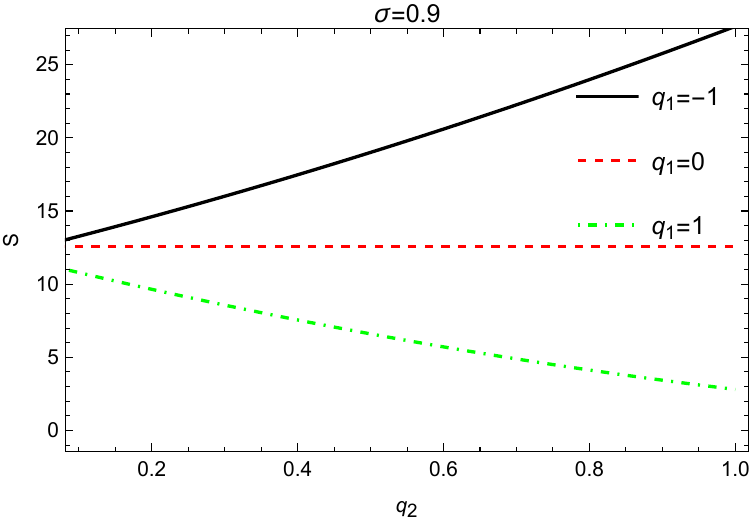}
\includegraphics[scale=0.7]{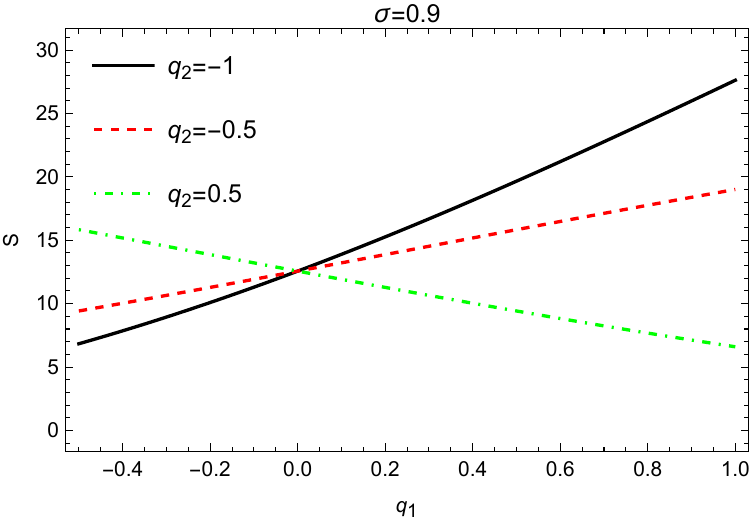}

\caption{ Entropy with respect $q_2$ for fixed values of $q_1$ (on the left) and with respect $q_1$ for fixed values of $q_2$ (on the right) with $\sigma=0.9$. }
 \label{Entropy}
    \end{center}
\end{figure*}

Enthalpy helps us to understand how energy changes in a black hole when it absorbs or emits energy.  Fig.~\ref{H} shows the radial dependence of black hole enthalpy in different $q_1$, $q_2$ and $\sigma$. The graph clearly shows that increasing the values of
$q_1$ and $q_2$ corresponds to higher enthalpy, which means higher internal energy. 

\subsection{Heat Capacity \label{sec3C}}

To check the thermodynamic stability of the compact objects, the heat capacity $C(r_h)$ of the objects is calculated. The positive (negative) specific heat signifies the local thermodynamic stability (instability) of the black holes. By using the relation 
\begin{equation}
C=\frac{\partial M_{BH}}{\partial T_{H}}=\frac{\partial M_{BH} / \partial r_{h}}{\partial T_{H} / \partial r_{h}}, 
\end{equation}
one finds the following expression for the heat capacity of Einstein Geometric-Proca compact object:
\begin{widetext}
\begin{equation}
\begin{split}
C=\frac{
2 \sqrt{3}  \pi r_h^2 \sqrt{( \sigma-3)}
\left(12 r_h^3 + l^2 \left(4 r_h + q_1^2 r_h^\sigma \sigma -q_1^2 r_h^\sigma \sigma^2  \right)\right)
}{432 r_h^6 (\sigma-3) 
+ 6 l^2 q_1 r_h^3 (1-\sigma) 
\left(2 q_2 \left( \sigma^2-3 - 2 \sigma\right) 
+ 3 q_1 r_h^\sigma \left(7 \sigma - 7 \sigma^2 + \sigma^3-1\right)\right) } 
\times\\ \times
\sqrt{
{
\left[12 r_h^3 +  4l^2 r_h + l^2q_1^2 r_h^\sigma \sigma(1-\sigma)\right]
\left[3 l^2 q_1^2 r_h^\sigma (1- 4\sigma+3\sigma^2)  
+ 2 (\sigma-3) 
\left(18 r_h^3 +6 r_hl^2 + q_1 q_2l^2 (\sigma^2-1)\right)\right]
}
}. 
\end{split}
\end{equation}
\end{widetext}
Fig. \ref{HC} shows the dependence of heat capacity on the radius of the horizon $r_h$. The graph indicates that the heat capacity of the Einstein-Geometric Proca AdS compact objects is negative and increases with increasing values of $q_1$. However, a change in $q_2$ results in a minor reduction in heat capacity. The negative value of the heat capacity indicates that the compact object is unstable in small radius.  However, for sufficiently large event horizon radius, the heat capacity becomes positive, indicating that the compact object is thermodynamically stable. As illustrated in Fig.~\ref{phase}, a phase transition occurs when the horizon radius $r_h$ becomes comparable to the AdS length scale $l$.     

\begin{figure*}[t]
    \begin{center}
    
    \includegraphics[scale=0.7]{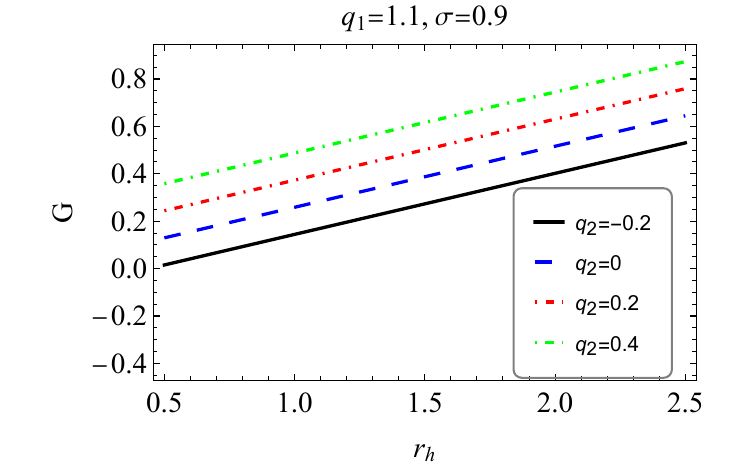}
    \includegraphics[scale=0.7]{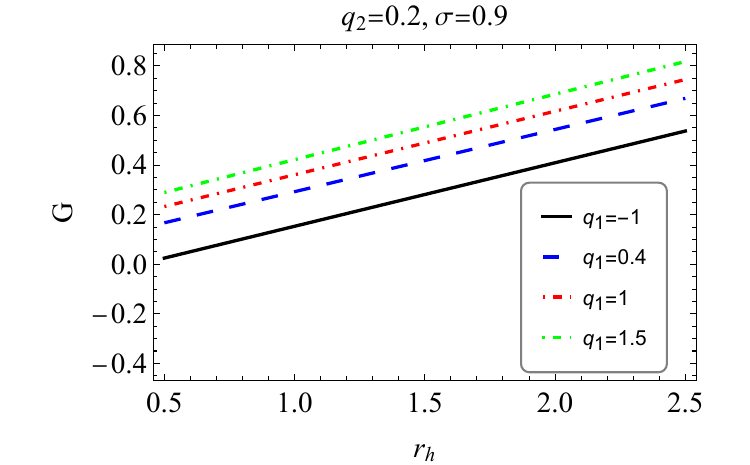}
   
     \caption{Gibbs energy with respect to horizon radius $r_h$ for fixed values of $q_1$, $q_2$ and $\sigma$.}
    \label{GE}
    \end{center}
\end{figure*}

\begin{figure*}[ht!]
 \begin{center}
   \includegraphics[scale=0.7]{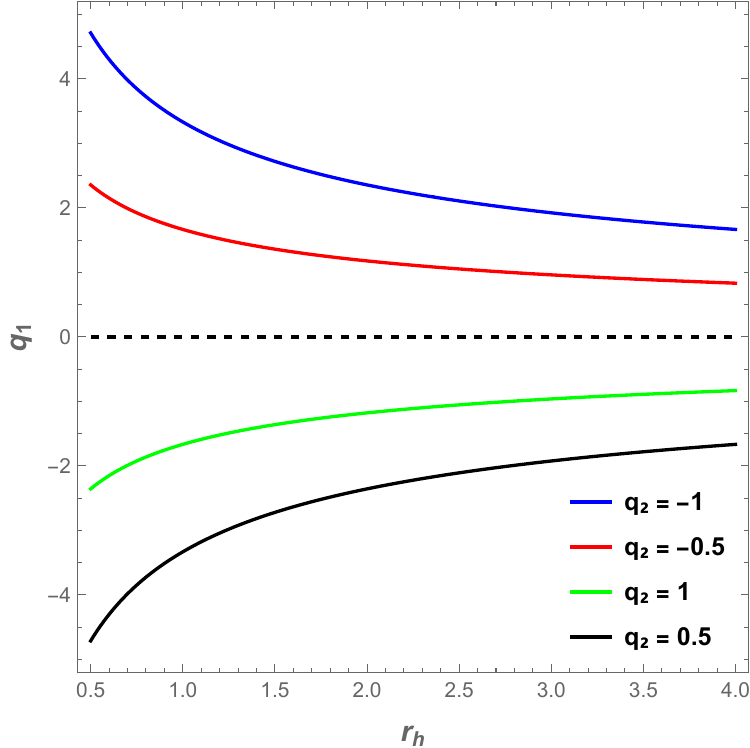}
   \caption{The contour plot of equation (\ref{delta}) for $\Delta=0$. }
\label{Firstlaw}
\end{center}
\end{figure*}

\subsection{Entropy \label{sec3D}}
The second law of thermodynamics implies that black holes must have entropy. Without it, adding mass to a black hole would violate this law. In 1972, Jacob Bekenstein suggested that a black hole's entropy is proportional to the area of its event horizon, which leads to the Hawking-Bekenstein entropy for a spherically symmetric compact object being defined by the area law \cite{Kastor2018}  
\begin{align}
S = \frac{A_h}{4 G_N},    
\end{align}
where $A_h = 4 \pi r_h^2$ is the area of the compact object horizon.
Fig. \ref{Entropy} shows how the black hole entropy varies with the parameters $q_1$ and $q_2$. The graphs reveal that these parameters significantly influence the entropy: when $q_1$ is negative, increasing $q_2$ increases the entropy, while for positive $q_1$, a higher $q_2$ leads to a reduced entropy. Similarly, for positive $q_2$, an increase in $q_1$ decreases the entropy, while for negative $q_2$, an increase in $q_1$ enhances it.
 
\subsection{Gibbs energy \label{sec3E}}

Gibbs free energy is a critical quantity in black hole thermodynamics, essential for analyzing both the thermodynamic stability and phase transitions of black holes. It serves as a fundamental indicator that helps determine the direction of thermodynamic processes in nature. The Gibbs free energy for a black hole is defined as follows \cite{Mahapatra2023}
$$G=M-TS.$$
In this context, $M, T, S$ represent the enthalpy, Hawking's temperature, and entropy of the black hole, respectively. Fig. \ref{GE} illustrates how the Gibbs free energy of the Einstein-Geometric Proca AdS compact object varies with the radial distance. The graphs clearly show that as the radial distance and the parameters $q_1$ and $q_2$ increase, the Gibbs free energy also increases.

\subsection{First Law of Thermodynamics}

Having established the expressions for entropy and temperature, we are now prepared to examine the first law of thermodynamics.
The first law of thermodynamics in our model is
\begin{equation}
dM = T\, dS + \Phi_2\, dq_2
\end{equation}
where
\begin{itemize}
    \item \( M = M(r_h) \): the mass of the compact object,
    \item \( T = T(r_h) \): the Hawking temperature,
    \item \( S = S(r_h) \): the entropy,
    \item \( \Phi_2  \): the potential of $q_2$ \, .
\end{itemize}
In the Maxwell limit ($\sigma \rightarrow 1$) the Proca field behaves as $\hat{\phi}(\hat{r}) \rightarrow q_1 + \frac{q_2}{\hat{r}}$ and it means that $q_2$ has the meaning of an electromagnetic-like charge while $q_1$ represents a uniform potential. Thus we dropped the term $ \Phi_1\, dq_1$ in the first law.
Since we consider the electric-type charge $q_2$ to be fixed in this variation, the first law reduces to its simple form:
\begin{equation}
\frac{dM}{dr_h} = T\, \frac{dS}{dr_h} \, .
\end{equation}
If define 
\begin{equation}
\Delta=
\frac{dM}{dr_h} - T\, \frac{dS}{dr_h}
\end{equation} 
we need to satisfy 
\begin{equation}
\Delta=0 \, .
\end{equation}
By substituting the explicit expressions for $M$, $T$ and $S$
into the first law, we obtain:
\begin{widetext}
\begin{align}
    \Delta&=\frac{1}{8} \left(\frac{12 r_h^2}{l^2}-q_1^2 (\sigma -1) \sigma  r_h^{\sigma -1}+4\right)-\left(\frac{r_h }{8 \sqrt{3}}\right)\nonumber\\
    &\times\sqrt{\frac{\left(l^2 \left(4 r_h-q_1^2 (\sigma -1) \sigma  r_h^{\sigma }\right)+12 r_h^3\right) \left(3 l^2 q_1^2 (\sigma -1) (3 \sigma -1) r_h^{\sigma }+2 (\sigma -3) \left(l^2 \left(q_1 q_2 \left(\sigma ^2-1\right)+6 r_h\right)+18 r_h^3\right)\right)}{l^4 r_h^4 (\sigma -3)}}
    \label{delta}
\end{align}   
\end{widetext}

To verify the validity of the first law, we plot the contour curves where \( \Delta = 0 \). Along each such curve in Fig. \ref{Firstlaw}, the first law of thermodynamics is exactly satisfied. The figure illustrates $(q_1, r_h)$ plane for various fixed values of $q_2$ in order to maintain the condition \( \Delta = 0 \). Notably, the case \( q_1 = 0 \), corresponding to the Schwarzschild-AdS solution, satisfies \( \Delta = 0 \). This special case is represented by the black dashed line in the plot.

\section{Conclusion \label{sec4}}

This study explores the thermodynamics of Einstein-Geometric Proca AdS compact objects, examining how Hawking radiation, entropy, enthalpy, heat capacity, and Gibbs energy depend on model parameters such as $q_1$, $q_2$ and $\sigma$. The study focuses on analyzing Hawking radiation, entropy, enthalpy, heat capacity, and Gibbs energy and their dependence of $q_1$, $q_2$, and $\sigma$.
Based on the calculations presented above, we arrive at the following conclusions:

\begin{itemize}
 \item We have analyzed the Hawking temperature of Einstein-Geometric Proca AdS compact objects and its dependence on $q_1$, $q_2$, and $r_h$ in Fig.\ref{HTr}. Our study reveals that temperature $T_H$ decreases monotonically with increasing radial distance. Moreover, higher values of $q_1$
are associated with higher temperatures near the compact object while higher values of $q_2$ corresponds to lower temperatures.

Fig.\ref{HT} shows that the Hawking temperature increases monotonically with both charge parameters while larger black hole masses correspond to lower temperatures. This behavior is consistent with the fact that heavier black holes are colder, whereas charge contributions enhance the surface gravity and raise the temperature.

 \item When examining the enthalpy of Einstein-Geometric Proca AdS compact objects, we observe that the enthalpy increases with the horizon radius (see Fig. \ref{H}.). Moreover, at a fixed
$\sigma$, adjustments in the parameters $q_1$ and $q_2$ lead to an increase in mass, thus increasing the enthalpy.

\item Additionally, we have examined the heat capacity, a key factor in determining the stability of the black hole in Fig. \ref{HC} . Our analysis indicates that the heat capacity of the Einstein-Geometric Proca AdS bcompact object is negative. Moreover, it increases with higher values of $q_1$, while increasing $q_2$ leads to a slight decrease.

\item The analysis of entropy for Einstein-Geometric Proca AdS black holes, as illustrated in Fig. \ref{Entropy}, shows its sensitivity to the parameters $q_1$ and $q_2$.  Specifically, for negative $q_1$ values, increasing $q_2$ leads to higher entropy, while for positive $q_1$ values, increasing $q_2$ causes the entropy to decrease. Similarly, if $q_2$ is positive, an increase in $q_1$ enhances the entropy, but if $q_2$ is negative, the entropy decreases as $q_1$ increases.

\item We have also examined the Gibbs free energy of Einstein-Geometric Proca AdS compact objects as shown in Fig. \ref{GE}. Our findings indicate that as the radial distance increases, along with the parameters $q_1$ and $q_2$, the Gibbs free energy also increases.
\end{itemize}

\section*{Acknowledgements}
The work of B.P. is supported by the Astrophysics Research Center of the Open University of Israel through The Israeli Ministry of Regional Cooperation. The work of B.P. is supported by Sabanc{\i} University, Faculty of Engineering and Natural Sciences. B.P. acknowledges the contribution of the COST Action CA21106 - COSMIC WISPers in the Dark Universe: Theory, astrophysics, and experiments (CosmicWISPers). This research is partially supported by the Research Grant F-FA-2021-510 of the Uzbekistan Ministry of Innovation Development.

\bibliographystyle{spphys}
\bibliography{reference,kerr-like}


\end{document}